\documentclass[aps,twocolumn,superscriptaddress,amsfont,amssymb,amsmath,showpacs,balancelastpage,english]{revtex4-1} 
\usepackage{babel}
\usepackage[dvipdfmx]{graphicx}
\usepackage{color}
\usepackage{natbib}
\usepackage{longtable}
\usepackage[utf8]{inputenc}

\definecolor{mygreen}{RGB}{0,130,0} 

\newcommand{\unit}[1]{\,\rm{#1}}


\begin{document}

\title{Neutron displacement noise-free interferometer for gravitational-wave detection}

\author{Atsushi Nishizawa}
\email{anishi@resceu.s.u-tokyo.ac.jp}
\affiliation{Research Center for the Early Universe (RESCEU), Graduate School of Science, The University of Tokyo, Tokyo 113-0033, Japan}

\author{Shoki Iwaguchi}
\affiliation{Department of Physics, Nagoya University, Furo-cho, Chikusa-ku, Nagoya, Aichi 464-8602, Japan}

\author{Yanbei Chen}
\affiliation{Theoretical Astrophysics 350-17, California Institute of Technology, Pasadena, California 91125, USA}

\author{Taigen Morimoto}
\affiliation{Department of Physics, Nagoya University, Furo-cho, Chikusa-ku, Nagoya, Aichi 464-8602, Japan}

\author{Tomohiro Ishikawa}
\affiliation{Department of Physics, Nagoya University, Furo-cho, Chikusa-ku, Nagoya, Aichi 464-8602, Japan}

\author{Bin Wu}
\affiliation{Department of Physics, Nagoya University, Furo-cho, Chikusa-ku, Nagoya, Aichi 464-8602, Japan}

\author{Izumi Watanabe}
\affiliation{Department of Physics, Nagoya University, Furo-cho, Chikusa-ku, Nagoya, Aichi 464-8602, Japan}

\author{Yuki Kawasaki}
\affiliation{Department of Physics, Nagoya University, Furo-cho, Chikusa-ku, Nagoya, Aichi 464-8602, Japan}

\author{Ryuma Shimizu}
\affiliation{Department of Physics, Nagoya University, Furo-cho, Chikusa-ku, Nagoya, Aichi 464-8602, Japan}

\author{Hirohiko Shimizu}
\affiliation{Department of Physics, Nagoya University, Furo-cho, Chikusa-ku, Nagoya, Aichi 464-8602, Japan}
\affiliation{Kobayashi-Maskawa Institute for the Origin of Particles and the
Universe, Nagoya University, Nagoya, Aichi 464-8602, Japan}

\author{Masaaki Kitaguchi}
\affiliation{Kobayashi-Maskawa Institute for the Origin of Particles and the
Universe, Nagoya University, Nagoya, Aichi 464-8602, Japan}
\affiliation{Department of Physics, Nagoya University, Furo-cho, Chikusa-ku, Nagoya, Aichi 464-8602, Japan}

\author{Yuta Michimura}
\affiliation{Department of Physics, University of Tokyo, Bunkyo, Tokyo 113-0033, Japan}

\author{Seiji Kawamura}
\affiliation{Department of Physics, Nagoya University, Furo-cho, Chikusa-ku, Nagoya, Aichi 464-8602, Japan}
\affiliation{Kobayashi-Maskawa Institute for the Origin of Particles and the
Universe, Nagoya University, Nagoya, Aichi 464-8602, Japan}

\begin{abstract}
An interferometer design that cancels all displacement noises of its test masses and maintains a gravitational-wave (GW) signal by combining multiple detector signals is called a displacement noise-free interferometer (DFI). The idea has been considered previously for a laser interferometer. However, a limitation of a laser DFI is that its sensitive frequency band is too high for astrophysical GW sources, $\sim 10^5\,{\rm Hz}$ even for a kilometer-sized interferometer. To circumvent this limitation, in this paper, we propose a neutron DFI, in which neutrons are used instead of light. Since neutrons have velocities much lower than the speed of light, the sensitive frequency band of a neutron DFI can be lowered down to $\sim 10^{-1}\,{\rm Hz}$. Therefore, a neutron DFI can be utilized for detecting GWs that are inaccessible by an ordinary laser interferometer on the ground. 
\end{abstract}

\date{\today}

\maketitle


\section{Introduction}

A gravitational wave (GW) was detected for the first time by LIGO and Virgo in 2015 \cite{GW150914:detection}, and the number of GW events observed so far is 90 until the end of the third observing run~\cite{LIGOScientific:2018mvr, LIGOScientific:2020ibl, LIGOScientific:2021usb,LIGOScientific:2021djp}. One type of noises that limits the sensitivity of a ground-based detector at lower frequencies is displacement noise, arising from seismic motions, thermal fluctuations of mirrors and suspensions, radiation pressure fluctuations of laser light. Displacement noise prevents us from improving detector sensitivity below $\sim 1\,{\rm Hz}$, e.g.~\cite{Advanced-GW-detectors:book, Dooley:2021vbi, Hall:2020dps}. Although there are rich targets of astrophysical GWs such as the early inspiral phase of stellar-mass binary black holes (BHs) and binary neutron stars and the coalescences of intermediate-mass BHs~\cite{Kawamura:2018esd,Kuns:2019upi}, a decihertz frequency band is inaccessible by ground-based detectors.

An idea to remove all displacement noises by combining multiple detector signals, while maintaining a GW signal, that is, a displacement noise-free interferometer (DFI) was originally proposed in \cite{Kawamura:2004cx, Chen:2005ya}. Later, more practical configurations have been considered in \cite{Chen:2006zra, Nishizawa:2008yr, Wang:2013lna}. The principle underlying DFIs was experimentally demonstrated with laser optics in \cite{Sato:2006gk, Kokeyama:2009ww}. In a series of the studies, it has been shown that a DFI can eliminate all displacement noises and improve sensitivity at lower frequencies, reaching $\Omega^2$ response to a GW where $\Omega$ is the angular frequency of a GW. However, a limitation of a laser DFI is that its sensitive frequency band is determined by the characteristic frequency of a detector, $f_{\rm c} \equiv c/(\pi L)$, where $L$ is the size of a detector and $c$ is the speed of light. Even for a kilometer-sized interferometer, the characteristic frequency is $\sim 10^5\,{\rm Hz}$, which is far above the signal frequencies of standard astrophysical GW sources expected.

To circumvent the limitation, in this paper, we propose to use matter waves, more specifically neutrons instead of light, in a DFI. A neutron interferometer~\cite{Rauch-Werner:book} was demonstrated on the basis of X-ray interferometer technique~\cite{Rauch:1974PhLA}. A silicon single crystal functions as a mirror for neutron waves according to Bragg's law. By machining four mirrors on a single block of a silicon crystal, the mirrors based on a crystal lattice are precisely aligned. Some remarkable experiments including studies of gravity were performed with the interferometer~\cite{Colella:1975dq}, however, the applications were limited by the block size. In addition, only monochromatic neutrons with the velocity of the order of 1\unit{km/s} could be applied to the interferometer because of the Bragg reflection. More recently, neutron interferometry using multilayer mirrors has been developed. By using two materials with different refractive indices for neutrons, mirrors and even beam splitters can be freely designed and produced. A multilayer mirror with gradually changing thickness enables us to utilize polychromatic neutrons for interferometry. The remote positioning of the artificial multilayers makes it possible to construct a large interferometer~\cite{Kitaguchi:2003PhRvA,Seki:2010JPSJ}. Since neutrons can have velocity much lower than the speed of light, the characteristic frequency $f_{\rm c}$ can be lowered down to the frequency band sensitive to astrophysical GW sources, $\sim 10^{-1}\,{\rm Hz}$, with the detector size of $\sim 3\unit{km}$. We use the unit of $c=1=\hbar$ throughout the paper.


\section{One-path signal}

We first derive a propagation equation for the phase of a massive particle in the presence of a GW. Then we integrate the differential equation and obtain a phase shift in a homogeneous gravitational field, that is, the Earth gravity.      

\subsection{Wave equation}

We start with the Klein-Gordon equation for a massive particle with mass $m$,
\begin{equation}
(\Box-m^2) \Phi =0 \;, 
\label{eq:KG-eq}
\end{equation}
where $\Box \equiv -\partial_t^2 + \nabla_i \nabla^i$. The wave function of a massive particle at a time $t$ and a position $\mathbf{X}$ at the zeroth order is taken as a plane wave, 
$\Phi_0 (t,\mathbf{X}) = A\, e^{-i(\omega t + \mathbf{k} \cdot \mathbf{X})}$. A perturbation for the plane wave in the presence of a GW is in general written as 
\begin{equation}
\Phi (t, \mathbf{X}) = \Phi_0 (t,\mathbf{X}) \left[ 1+ i \phi_{\rm gw} (t,\mathbf{X}) \right] \;. \\
\end{equation}
We assume that the angular frequency of a GW, $\Omega$, is much smaller than that of a massive particle, namely, $\Omega \ll \omega$ and that a massive particle is nonrelativistic, $k \ll \omega$. Hereafter we approximately use $m$ instead of $\omega$.
 
For a GW propagating in a flat spacetime in the transverse-traceless gauge, we write
\begin{equation}
ds^2 =-dt^2 + [\eta_{ij}+h_{ij}(t,\mathbf{X})] dx^i dx^j \;.
\label{eq:gw-z}
\end{equation}
In this coordinate system, the wave equation (\ref{eq:KG-eq}) at the leading order in $h$ is reduced to    
\begin{equation}
\frac{\partial \phi_{\rm gw}}{\partial t}  = -\frac{h^{ij} k_i k_j}{2\omega} \approx -\frac{h^{ij} k_i k_j}{2m} \;. 
\label{eq:prop-eq}
\end{equation}

\begin{figure}[t]
\begin{center}
\includegraphics[width=7cm]{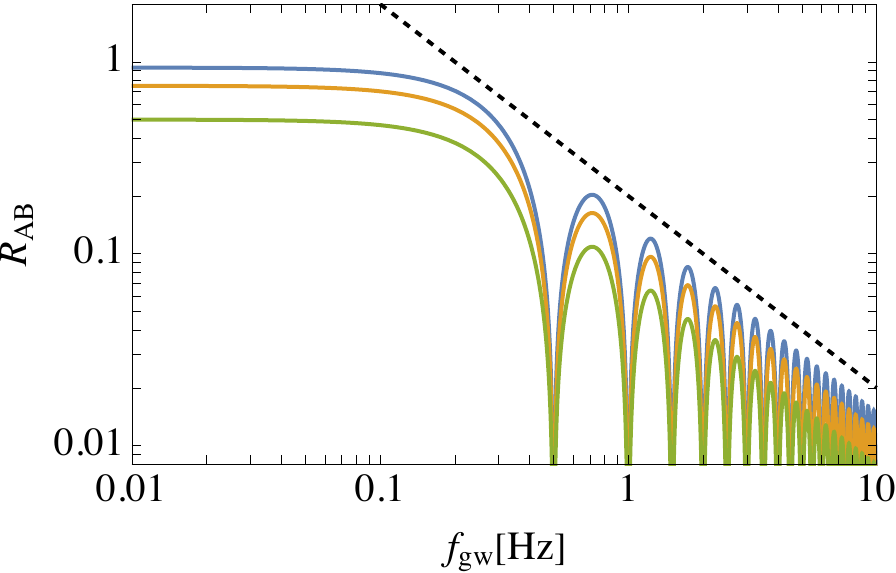}
\includegraphics[width=7cm]{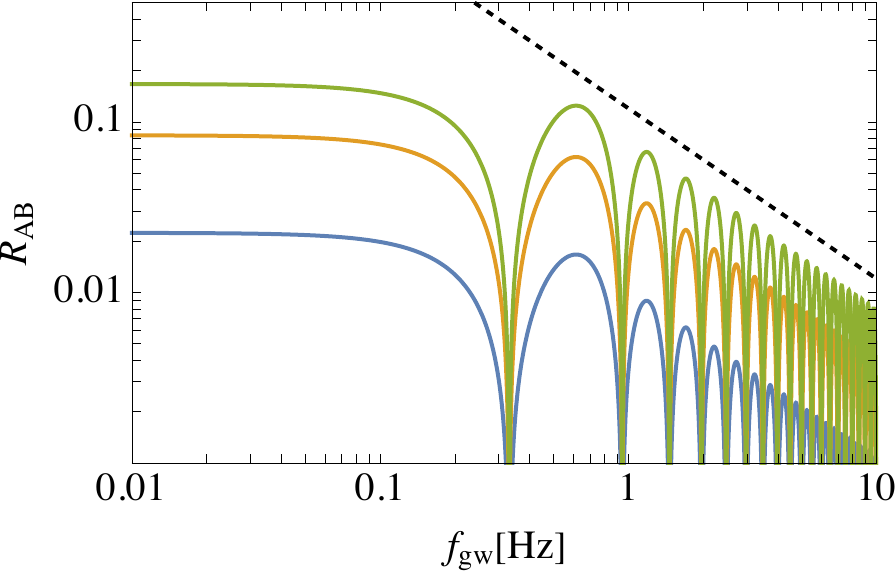}
\caption{GW responses for one-way paths with neutron initial angles, $\alpha=\pi/24$ (blue), $\pi/12$ (orange), and $\pi/6$ (green). The initial velocities are chosen as $v_0=75.1$, $37.9$, and $19.6\,{\rm m}/{\rm s}$, respectively, fixing the flight time of neutrons to $2\,{\rm s}$. Upper and lower panels are the responses to a GW propagating in the $z$ and $x$ directions, respectively. The diagonal dashed lines are proportional to $f_{\rm gw}^{-1}$.}
\label{single-path}
\end{center}
\end{figure}

\subsection{One-path signal on the Earth}

Suppose that a gravitational field (due to the Earth) is uniform in the $z$ direction and that a massive particle is emitted with velocity $v_0$ along the direction of angles ($\alpha$, $\beta$) at the time $t=0$. By denoting $\mathbf{X}(t)=\mathbf{X}_0+\mathbf{x}(t)$, the trajectory of the particle is 
\begin{equation}
\mathbf{x}(t) = \left\{ v_0 t \cos \alpha \cos \beta \;, v_0 t \cos \alpha \sin \beta \;, v_0 t \sin \alpha -\frac{g}{2} t^2 \right\} \;,
\end{equation}
with the end point given by setting $t=T$. The wave-number vector is
\begin{equation}
\mathbf{k}(t) = m v_0 \left\{ \cos \alpha \cos \beta, \cos \alpha \sin \beta, \sin \alpha - f_0 t \right\} \;. 
\label{eq:k-vector}
\end{equation}
where $f_0 \equiv g/v_0$. Integrating Eq.~(\ref{eq:prop-eq}) from the point A at $t$ to the point B at $t+T$ gives a phase shift due to the GW
\begin{equation}
\phi^{\rm gw}_{\rm AB}(t) \approx -\frac{k_0^2}{2m} \int_t^{t+T} h^{ij}\left[t^{\prime}, \mathbf{X}(t^{\prime})\right] \tilde{k}_i (t^{\prime}) \tilde{k}_j (t^{\prime}) dt^{\prime} \;.
\label{eq:1-link}
\end{equation}
where we defined the dimensionless wave number $\tilde{k}_i \equiv k_i/k_0$ with $k_0 \equiv mv_0$. We expanded the GW amplitude at the position $\mathbf{X}(t)$ as $h^{ij}(t,\mathbf{X}(t)) \approx h^{ij}(t,\mathbf{X}_0) + \mathbf{x}(t) \cdot \mathbf{\nabla} h^{ij}(t,\mathbf{X}_0)$, which gives a factor, $1+ i \mathbf{k}_{\rm gw} \cdot \mathbf{x}(t)$. However, the second term can be ignored because the typical size of interferometers that we consider is $x \sim 3\unit{km}$ and  $k_{\rm gw} x \sim x/\lambda_{\rm gw}$ is negligibly small for GWs at $1\unit{Hz}$, whose wavelength is $\lambda_{\rm gw}$ is $3 \times 10^5\unit{km}$. For simplicity, we can set $\mathbf{X}_0$ to zero without loss of generality and drop the position dependence from $\phi^{\rm gw}_{\rm AB}$.

In practice, the timing of neutron detection at the points A and B and the displacement of the detectors introduce noise in the GW signal. We call these noises clock and displacement noises of detectors at A and B (more concretely those of mirrors and beam splitters in the later section), which are given by
\begin{align}
\phi^{\rm clock}_{\rm AB} (t) &\approx m \{ \tau_{\rm  B} (t+T) - \tau_{\rm  A} (t) \} \;, \\
 \phi^{\rm mirror}_{\rm AB} (t) &= 2 \mathbf{k}^{\rm AB}_{\rm  B} \cdot \mathbf{d}_{\rm  B} (t+T) - 2 \mathbf{k}^{\rm AB}_{\rm  A} \cdot \mathbf{d}_{\rm A} (t) \;, 
 \label{eq:displacement} 
\end{align}
where $\tau_{\rm A}$ and $\tau_{\rm B}$ are the clock noises of the detectors A and B, and $\mathbf{d}_{\rm A}$ and $\mathbf{d}_{\rm B}$ are the displacement noise vectors of the detectors A and B, respectively, and $\mathbf{k}^{\rm AB}_{\rm A}$ is the wave-number vector at the point A on the path from A to B and so on. 

Defining the Fourier transform of the GW amplitude by
\begin{equation}
H^{ij}(\Omega) \equiv \int_{-\infty}^{\infty} dt\, e^{i\Omega t} h^{ij} (t) \;,
\end{equation}
and from Eq.~(\ref{eq:1-link}), we obtain the single-path signal in the Fourier domain
\begin{align}
\Phi_{\rm AB} (\Omega) &= -\frac{k_0^2}{2m} \biggl\{ P_0 (\Omega) \tilde{k}_I \tilde{k}_J H^{IJ} (\Omega)  \nonumber \\
&+ \left( \sin \alpha \, P_0 (\Omega) + P_1 (\Omega) \right) \tilde{k}_I H^{Iz} (\Omega) \nonumber \\
& + \left( \sin^2 \alpha \, P_0 (\Omega) + 2 \sin \alpha \, P_1 (\Omega) + P_2 (\Omega) \right) H^{zz} (\Omega) \biggr\} \nonumber \\
& + m \left\{ w(\Omega) \tau_{\rm  B} (\Omega) - \tau_{\rm A} (\Omega) \right\} \nonumber \\
&+ 2 w(\Omega) \mathbf{k}^{\rm AB}_{\rm B} \cdot \mathbf{d}_{\rm  B} (\Omega) - 2 \mathbf{k}^{\rm AB}_{\rm A} \cdot \mathbf{d}_{\rm A} (\Omega) \;,
\label{eq:sig-1link}
\end{align}
where the indices are $I, J=x,y$. We defined the phase $w(\Omega)=e^{-i \Omega T}$ and the phase responses 
\begin{align}
P_0 (\Omega) &\equiv -\frac{i}{\Omega} \left\{ 1-w(\Omega) \right\} \;, \\
P_1 (\Omega) &\equiv \frac{f_0}{\Omega^2} \left\{1- w(\Omega) \left( 1+ i \Omega T \right) \right\} \;, \\
P_2 (\Omega) &\equiv \frac{2i f_0^2}{\Omega^3} \left\{1- w(\Omega) \left( 1+ i \Omega T -\frac{1}{2} \Omega^2 T^2 \right) \right\} \;.
\end{align}
To study the performance of a neutron DFI, it is convenient to define a GW response function normalized by the GW amplitude, $H$, and the parameter combination of a neutron interferometer, $\eta$, 
\begin{equation}
R_{\rm AB} (\Omega) \equiv \frac{\Phi^{\rm gw}_{\rm AB} (\Omega)}{\eta H} \;, \quad \eta \equiv \frac{k_0^2 T}{2m} \;.
\label{eq:res-1link}
\end{equation} 

Let us first look at a GW response in a simple case, that is, a parabolic trajectory of a neutron on the $x$-$z$ plane ($\beta=0$). For a GW propagating in the $z$ direction ($\theta=0,\phi=0$) with $\psi=0$, the GW response function from Eq.~(\ref{eq:res-1link}) is
\begin{equation}
R_{\rm AB}(\Omega) = \cos^2 \alpha \, \left| \frac{\sin(\Omega T/2)}{\Omega T/2} \right| \;.
\end{equation}
For a GW propagating in the $x$ direction ($\theta=\pi/2,\phi=0$) with $\psi=0$, the GW response function is
\begin{equation}
R_{\rm AB}(\Omega) = \frac{1}{T} \left| \sin^2 \alpha \, P_0 (\Omega) + 2 \sin \alpha \, P_1 (\Omega) + P_2 (\Omega) \right| \;.
\end{equation}
In Fig.~\ref{single-path}, the response functions to a GW propagating in specific directions are plotted for neutron trajectories with different $\alpha$ by fixing the flight time to $T=1\,{\rm s}$. When a GW comes from an arbitrary direction, $(\theta, \phi)$, with the polarization angle $\psi$, the GW components are computed by
\begin{align}
h_{ij}^{\prime} = {\cal R}_{ia} {\cal R}_{jb} h_{ab} = ({\cal R} h {\cal R}^{\rm T})_{ij} \;,
\end{align}
with the rotation matrix ${\cal R} = {\cal R}_z(-\phi) {\cal R}_y(\theta) {\cal R}_z(-\psi)$ where ${\cal R}_y(\theta)$ rotates the propagating direction of a GW about the y axis by $\theta$ and so on.

\begin{figure}[t]
\begin{center}
\includegraphics[width=7.5cm]{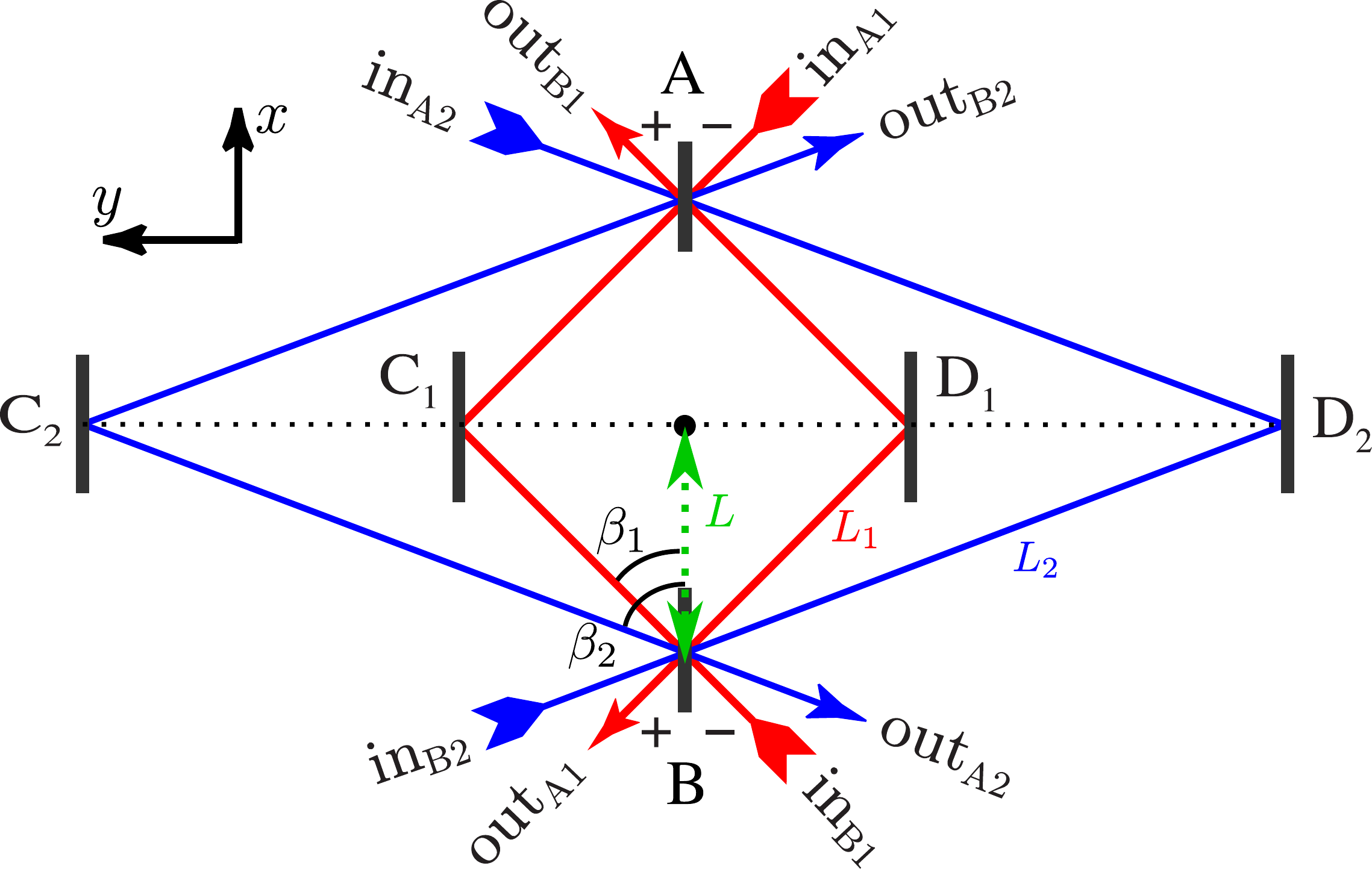}
\caption{Neutron-DFI configuration. The signs near the beam splitters, A and B, represent the sign convention for reflectivities.}
\label{fig:DFI-design}
\end{center}
\end{figure}

\section{DFI signal combination}

The DFI configuration we consider is shown in Fig.~\ref{fig:DFI-design}. The mirrors $C_1$, $C_2$, $D_1$, and $D_2$ and the beam splitters $A$ and $B$ are arranged on the x-y plane, but the DFI configuration is three dimensional, including the $z$ direction because neutrons propagate along parabolic paths. For the two different paths 1 and 2 in Fig.~\ref{fig:DFI-design}, we inject neutrons with the same velocity, $v_0$. They are bounced at $C_1$, $C_2$, $D_1$, and $D_2$ without losing speeds. In the Fourier domain, the DFI signal combination that cancels all displacement noises is 
\begin{align}
\Phi_{\rm DFI}(\Omega) &= \frac{1}{2\Omega \bar{T}} \bigl[ c_1(\Omega) \left\{ \Phi_{\rm A1B}(\Omega) -  \Phi_{\rm B1A}(\Omega) \right\} \nonumber \\
& \qquad \;\; - c_2(\Omega) \left\{ \Phi_{\rm A2B}(\Omega) - \Phi_{\rm B2A}(\Omega) \right\} \bigr] \;,
\label{eq:DFI2D-f} \\
\Phi_{{\rm A}i{\rm B}}(\Omega) &\equiv \Phi_{{\rm AC}_i}(\Omega) + w_i(\Omega) \Phi_{{\rm C}_i{\rm B}}(\Omega) \nonumber \\
&- \Phi_{{\rm AD}_i}(\Omega) - w_i(\Omega) \Phi_{{\rm D}_i{\rm B}}(\Omega) \;, 
\end{align}
with the coefficients
\begin{equation}
c_1(\Omega) = \frac{1-w_2^2(\Omega)}{\cos \alpha_1 \sin \beta_1}\;, \quad c_2(\Omega) = \frac{1-w_1^2(\Omega)}{\cos \alpha_2 \sin \beta_2}  \;, 
\end{equation}
where $i=1,2$, $T_i$ is the flight time of each individual parabolic trajectory, $w_i(\Omega)=e^{-i \Omega T_i}$ is the phase shift corresponding to $T_i$, and the average of the flight time is $\bar{T} \equiv (T_1+T_2)/2$. The combinations, $\phi_{\rm A1B} -\phi_{\rm B1A}$ and $\phi_{\rm A2B} -\phi_{\rm B2A}$, cancel the displacement noises of mirrors $C_1$, $C_2$, $D_1$, and $D_2$. The clock and displacement noises at the beam splitters remain in $\phi_{\rm A1B} -\phi_{\rm B1A}$ and $\phi_{\rm A2B} -\phi_{\rm B2A}$, but they can be eliminated by taking a linear combination with the appropriate coefficients. The coefficients, $c_1$ and $c_2$, are divided by a factor of $2\Omega \bar{T}$ so that the coefficients give frequency-independent constants at low frequencies and the original frequency dependence of the DFI response is preserved. Therefore, in the DFI signal, all displacement noises are canceled out but a GW signal remains.

Given the parameters, $L$, $\beta_1$, $\beta_2$, and $v_0$, the other parameters of the DFI configuration are uniquely fixed to
\begin{equation}
L_i = \frac{L}{\cos \beta_i} \;, \quad \sin 2 \alpha_i = \frac{g L_i}{v_{0}^2} \;, \quad T_i = \frac{L_i}{v_{0} \cos \alpha_i} \;, \label{eq:length-angle-relation}
\end{equation}
where $i=1,2$ and $T_i$ is the flight time of neutrons, that is, the time when a neutron returns at the height of $z=0$. For $L=3\,{\rm km}$, $\beta_1=\pi/4\,{\rm rad}$, $\beta_2=\pi/3\,{\rm rad}$, and $v_0 =3\,{\rm km}/{\rm s}$, the derived parameters are $T_1=\sqrt{2}\,{\rm s}$, $T_2=2\,{\rm s}$, $\alpha_1=0.132\,{\rm deg}$, and $\alpha_2=0.187\,{\rm deg}$. This set of parameters is not optimal, but we take it as fiducial parameters to illustrate the DFI response to a GW.

\begin{figure}[t]
\begin{center}
\includegraphics[width=7cm]{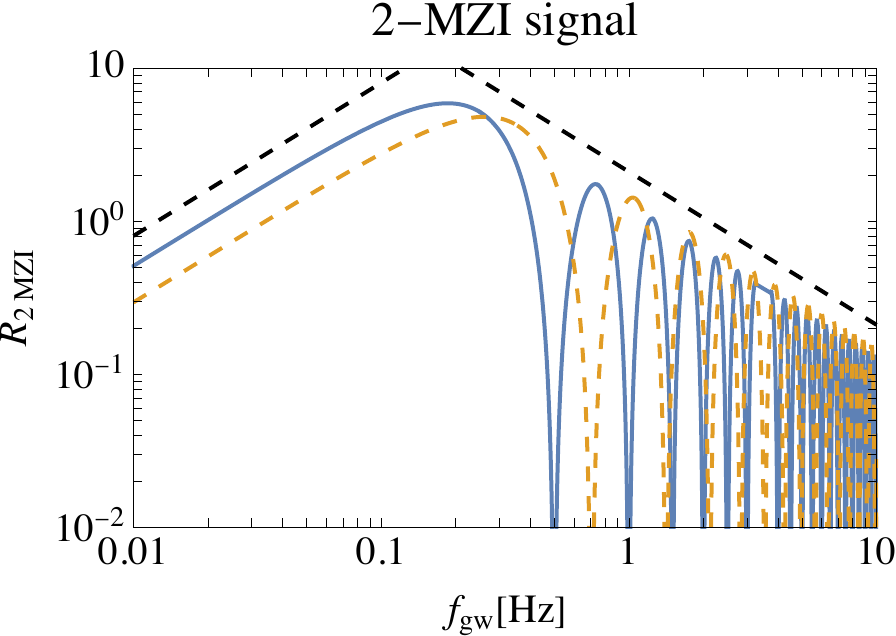}
\includegraphics[width=7cm]{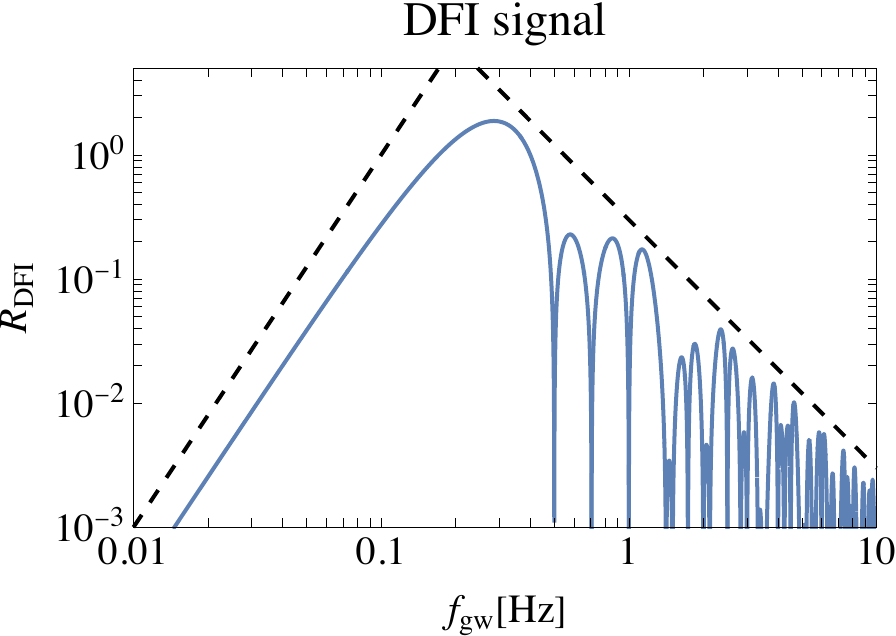}
\caption{DFI response function for a GW propagating from $+z$ direction as a function of GW frequency, $f_{\rm gw}=\Omega/(2\pi)$. The polarization angle is set to $\psi=\pi/4$. Top: the response of a bidirectional MZI, $R_{\rm 2MZI}(\Omega)$, for $\Phi_{\rm A1B} - \Phi_{\rm B1A}$ (orange, dashed) and $\Phi_{\rm A2B} - \Phi_{\rm B2A}$ (blue, solid). The diagonal dashed lines are just for reference and are proportional to $f_{\rm gw}^{-1}$. Bottom: the DFI response, $R_{\rm DFI}(\Omega)$, for Eq.~(\ref{eq:DFI2D-f}) The diagonal dashed lines are proportional to $f_{\rm gw}^{3}$ and $f_{\rm gw}^{-2}$, respectively. 
}
\label{fig:DFI-response}
\end{center}
\end{figure}

Figure~\ref{fig:DFI-response} shows the DFI response, 
\begin{equation}
R_{\rm DFI} (\Omega) \equiv \frac{|\Phi_{\rm DFI} (\Omega)|}{\bar{\eta} H} \;, \quad \bar{\eta} = \frac{k_0^2 \bar{T}}{2m} \;.
\end{equation}
By taking the combinations of a bidirectional Mach-Zender interferometer (MZI), each corresponding to the first and second lines in Eq.~(\ref{eq:DFI2D-f}), all the displacement noises of mirrors $C_1$, $C_2$, $D_1$, and $D_2$ are canceled out, while a GW response is the same as a single MZI, that is, proportional to $\Omega$ at low frequencies. Further, by taking the linear combination of two such MZIs with the appropriate coefficients for the full DFI signal, the displacement noises at the beam splitters are also canceled out. A GW signal is partially canceled but still remains, with a frequency dependence of the GW response proportional to $\Omega^3$ at low frequencies. This is the GW response consistent with that of the 2-dimensional configuration of a laser DFI \cite{Chen:2006zra}. However, a remarkable difference is that the characteristic frequency of the GW response, $f_{\rm c} = v_0/(\pi L) \approx 0.32\unit{Hz}$, is significantly lowered down below $\sim 1\unit{Hz}$ because of slow velocities of neutrons. The GW responses of the bidirectional MZIs have dips at the frequencies of multiples of $T_1^{-1}$ or $T_2^{-1}$, which are $0.71\,{\rm Hz}$ and $0.50\,{\rm Hz}$ and at which GW signals vanish. In the DFI signal, dips are much narrower because they appear only at the frequencies that the GW signals of two bidirectional MZIs are canceled out exactly.

At the low frequency limit, that is, $\Omega \ll \Omega_{\rm c} \equiv 2\pi f_{\rm c}$, the DFI response in Eq.~(\ref{eq:DFI2D-f}) is expanded in powers of $\Omega$ and the leading-order term is
\begin{align}
R_{\rm DFI}(\Omega) &= \frac{64}{3} \left(\frac{\Omega}{\Omega_{\rm c}}\right)^3 {\cal G}_{\rm s}(\theta, \phi, \psi) {\cal G}_{\rm d}(\alpha_1, \alpha_2, \beta_1, \beta_2) \;, \\
{\cal G}_{\rm s} & \equiv \frac{1+\cos^2 \theta}{2} \sin 2\phi \cos 2\psi \nonumber \\
\qquad & \;+ \cos \theta \cos 2 \phi \sin 2\psi \;, \\
{\cal G}_{\rm d} &\equiv - \frac{1}{\cos \alpha_1 \cos \alpha_2 \cos \beta_1 \cos \beta_2} \nonumber \\
\qquad & \; \times \frac{\cos \alpha_1 \cos \beta_1-\cos \alpha_2 \cos \beta_2}{\cos \alpha_1 \cos \beta_1+\cos \alpha_2 \cos \beta_2} \;. 
\label{eq:geometrical-factor-d}
\end{align}
The factors, ${\cal G}_{\rm s}$ and ${\cal G}_{\rm d}$, are the geometrical factors determined by the angle parameters for a source direction and a detector, respectively. In addition to the low-frequency suppression of the GW response, $(\Omega/\Omega_{\rm c})^3$, Eq.~(\ref{eq:geometrical-factor-d}) can be expressed in terms of flight times by using Eq.~(\ref{eq:length-angle-relation}) and gives a suppression factor proportional to the flight time difference:
\begin{equation}
{\cal G}_{\rm d} = \frac{1}{2\cos \alpha_1 \cos \alpha_2 \cos \beta_1 \cos \beta_2} \frac{T_2 - T_1}{\bar{T}} \;. 
\end{equation}
However, the suppression is modest ($|T_2 - T_1|/\bar{T} \approx 0.34$) for the parameter set above. Interestingly, the dependence on a GW source direction is factored out and is given by the same expression as that of a Michelson-type GW detector. 
The optimal direction of a GW response is the $z$ axis. The angular average over a GW source direction, $(\theta, \phi)$, and a polarization angle, $\psi$, decreases the GW response by a factor of $2.24$ but does not change the spectral shape of the GW response.

\section{Sensitivity}

To implement the neutron DFI in reality, experimental feasibility should be evaluated. Since a displacement noise of the neutron DFI is already canceled, we estimate a shot noise arising from the finite number of neutrons in the output ports. 

From Eq.~(\ref{eq:DFI2D-f}), a GW signal in the DFI combination is given by
\begin{align}
\Phi_{\rm DFI}^{\rm gw}(\Omega) &= \frac{1}{2\Omega \bar{T}} \left[ c_1(\Omega) \left\{ \Phi_{\rm A1B}^{\rm gw}(\Omega) -  \Phi_{\rm B1A}^{\rm gw}(\Omega) \right\} \right. \nonumber \\
& \qquad \quad \left. - c_2(\Omega) \left\{ \Phi_{\rm A2B}^{\rm gw}(\Omega) - \Phi_{\rm B2A}^{\rm gw}(\Omega) \right\} \right] \nonumber \\
&= \bar{\eta} R_{\rm DFI} (\Omega) H \;.
\label{eq:signal-DFI-Fourier} 
\end{align}
In the GW detection experiments, a GW signal is conventionally given in the unit of $\unit{Hz}^{-1/2}$ as 
\begin{equation}
S(\Omega) \equiv \left| \Phi_{\rm DFI}^{\rm gw}(\Omega) \right| \Omega^{1/2} \;.
\label{eq:-signal-DFI-rHz}
\end{equation}
While defining the power spectral density of shot noise by 
\begin{equation}
\langle \Phi_{\rm DFI}^{\rm shot}(\Omega) \Phi_{\rm DFI}^{\rm shot}(\Omega^{\prime}) \rangle = 2\pi \delta (\Omega - \Omega^{\prime}) P_{\rm n}(\Omega)\;,
\end{equation}
the shot noise for the DFI combination is 
\begin{align}
N^2 (\Omega) &\equiv P_{\rm n}(\Omega) \nonumber \\
&= \frac{1}{(2\Omega \bar{T})^2} \left[ |c_1(\Omega)|^2 \left\{ P^{\rm shot}_{\rm A1B}(\Omega)  + P^{\rm shot}_{\rm B1A}(\Omega) \right\} \right. \nonumber \\
& \qquad \qquad \left. + |c_2(\Omega)|^2 \left\{ P^{\rm shot}_{\rm A2B}(\Omega)  + P^{\rm shot}_{\rm B2A}(\Omega)  \right\} \right] \nonumber \\
&= \frac{1}{2\Omega^2 \bar{T}^2} \left\{ |c_1(\Omega)|^2 + |c_2(\Omega)|^2 \right\} P_1^{\rm shot}(\Omega).
\label{eq:noise2-DFI-PSD}
\end{align}
At the last line, we assume that four MZIs are identical and have same shot noise amplitude, that is, $P_1^{\rm shot}=P^{\rm shot}_{\rm A1B} = P^{\rm shot}_{\rm B1A}=P^{\rm shot}_{\rm A2B} = P^{\rm shot}_{\rm B2A}$. Given a neutron flux $F$, the shot noise of a single MZI is given by the Poissonian fluctuation of neutron flux as $P_1^{\rm shot}(\Omega) = 1/F$. Substituting this for Eq.~(\ref{eq:noise2-DFI-PSD}) and setting $S/N=1$, we obtain the shot-noise-limited sensitivity to GW amplitude in $\unit{Hz}^{-1/2}$ 
\begin{align}
h_n (\Omega) &\equiv H_{\rm n}(\Omega) \Omega^{1/2} \nonumber \\
& = \frac{1}{\bar{\eta} R_{\rm DFI} (\Omega)} \sqrt{\frac{|c_1(\Omega)|^2 + |c_2(\Omega)|^2}{2\Omega^2 \bar{T}^2 F}} \;.
\label{eq:sens-general}
\end{align}

\begin{figure}[t]
\begin{center}
\includegraphics[width=8cm]{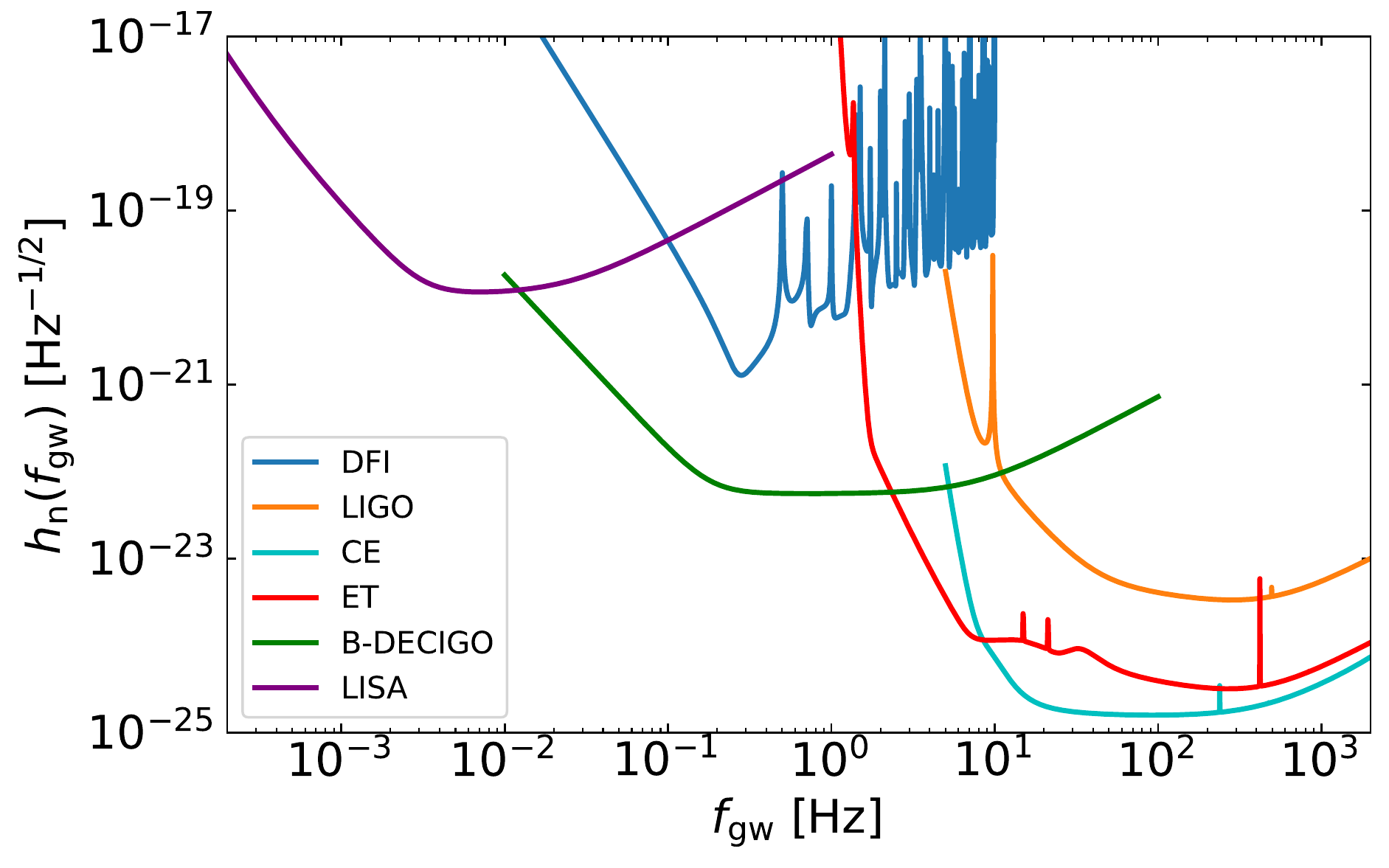}
\caption{Sky-averaged shot-noise-limited sensitivity to GWs. The curves plotted are the noise curves of the neutron DFI (this work, blue), LIGO \cite{LIGOScientific:2014pky} (orange), Einstein Telescope (ET) \cite{Maggiore:2019uih} (red), Cosmic Explorer (CE) \cite{Evans:2016mbw} (cyan), B-DECIGO \cite{Sato:2017dkf} (green), and LISA \cite{Robson:2018ifk} (purple).}
\label{fig:DFI-sens}
\end{center}
\end{figure} 

The largest flux of currently available neutron beams is that produced in J-PARC \cite{Maekawa:2010NIMPA} or ESS planned in the near future \cite{Garoby:2018PhyS}, $F\sim 10^{13}$ neutrons$/$s at the wavelength $\lambda_{\rm n}=0.13\unit{nm}$, corresponding to $v_0=3\unit{km/s}$. If the flux is injected in the neutron DFI with $L=3\unit{km}$, $\beta_1=\pi/4\,{\rm rad}$, and $\beta_2=\pi/3\,{\rm rad}$, the shot-noise-limited sensitivity at the characteristic frequency, $f_{\rm c} = 0.32\unit{Hz}$ for the above parameters, is written as
\begin{align}
h_{\rm n} (f_{\rm c}) & = 6.8 \times 10^{-22} \left( \frac{3\unit{km/s}}{v_0} \right) \left( \frac{3\unit{km}}{L} \right) \left( \frac{1.78}{R_{\rm DFI} (f_{\rm c})} \right) \nonumber \\
& \quad \times \left( \frac{10^{13}\unit{s}^{-1}}{F} \right)^{1/2} \unit{Hz}^{-1/2} \;.
\end{align}
In Fig.~\ref{fig:DFI-sens}, we plot the noise curves of the neutron DFI and those of future detectors except for LIGO for reference. It is explicitly shown that the neutron DFI is sensitive to the lower frequency band that ground-based detectors are inaccessible. Furthermore, the neutron DFI  bridges the frequency bands of LISA and ground-based laser interferometers and plays a crucial role for multi-band observations because the mission life time of the neutron DFI can be much longer than that of space-based missions.

Next we estimate the horizon redshift $z_{\rm max}$ and the corresponding luminosity distance $d_{\rm L}$ for equal-mass binary BHs as a function of total mass, $m_1+m_2$ (equal mass), in a source frame, based on the sky-averaged noise curve of a neutron DFI. The signal-to-noise ratio (SNR) $\rho$ is computed from
\begin{equation}
\rho^2 = 4 \int_{f_{\rm{min}}}^{f_{\rm{max}}} \frac{|H (f)|^2}{h_{\rm n}^2(f)} df \;.
\label{eq:SNR}
\end{equation}
The lower cutoff frequency $f_{\rm min}$ of the integration is determined by comparing the low frequency cutoff of detector sensitivity and the frequency at which the observational time is equal to the time to merger. In the case of a neutron DFI, the low frequency cutoff of the detector sensitivity is always higher in the range of BH mass that we consider, $10\,M_{\odot}$ and $10^6\,M_{\odot}$, and for the observation time of 1\,year. We set it to $f_{\rm min}=0.02\unit{Hz}$. While the higher frequency cutoff is set to the higher one between the highest frequency of a GW waveform or the high frequency cutoff of the detector sensitivity, $10\unit{Hz}$ for a neutron DFI. For the waveform, we will use the inspiral-merger-ringdown phenomenological (IMR PhenomD) waveform~\cite{Khan:2016PRD} (compiled in Appendix of \cite{Nishizawa:2017nef}), which is an waveform for aligned-spinning (nonprecessing) binary BHs.

\begin{figure}[t]
\begin{center}
\includegraphics[width=8.5cm]{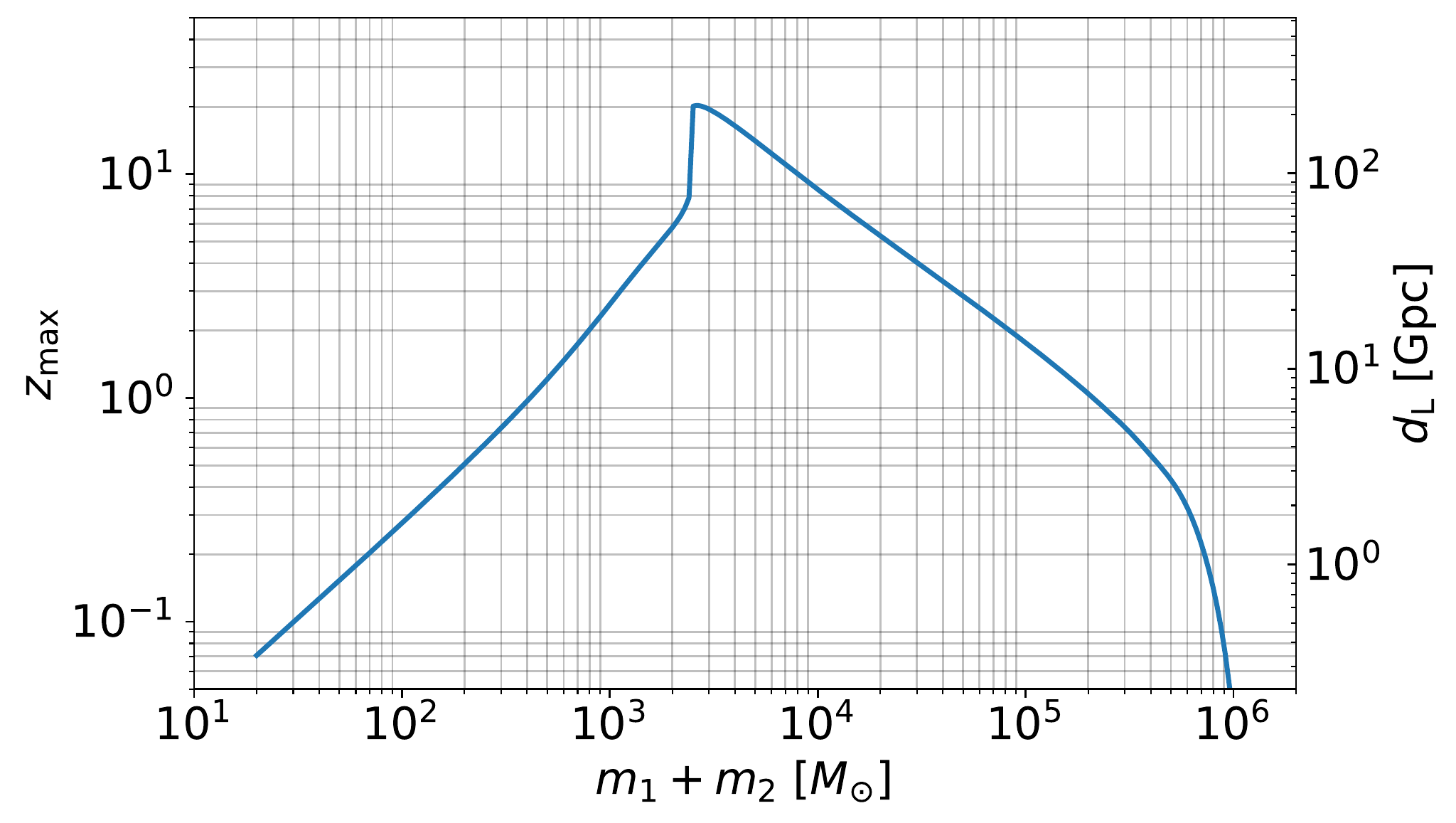}
\caption{Sky-averaged horizon redshift $z_{\rm max}$ (luminosity distance $d_{\rm L}$) for equal-mass binary BHs as a function of total mass, $m_1+m_2$, in a source frame.}
\label{fig:DFI-BBH-horizon}
\end{center}
\end{figure}

In Fig.~\ref{fig:DFI-BBH-horizon}, the horizon redshift $z_{\rm max}$ and the corresponding luminosity distance $d_{\rm L}$, defined by the SNR $\rho=8$, for equal-mass binary BHs are shown as a function of total mass, $m_1+m_2$, in a source frame. The neutron DFI can detect binary BHs with the broad range of masses, from stellar-mass BHs to supermassive BHs. Particularly, the mass range above $\sim 10^2\,M_{\odot}$ cannot be accessed from the ground-based laser interferometers such as LIGO, ET, and CE. While as expected from the low-mass end in Fig.~\ref{fig:DFI-BBH-horizon}, binary neutron stars and binaries of neutron-star and BH are difficult to observe because of smaller GW amplitude. For $30\,M_{\odot}$-$30\,M_{\odot}$ binary BHs, the horizon redshift of a neutron DFI is $z_{\rm max} \sim 0.18$. Based on the median of the merger rate of stellar-mass BBHs constrained by LIGO and Virgo, $23.9\,{\rm Gpc}^{-3}\,{\rm yr}^{-1}$ \cite{LIGOScientific:2020kqk}, the number of detections expected with a neutron DFI is $\sim 37$ per year. For binary BHs with total masses, $10^3\,M_{\odot}$, $10^4\,M_{\odot}$, and $10^5\,M_{\odot}$, the horizon redshifts are $z_{\rm max} \sim  2.6$, $8.5$, and $1.8$, respectively. Since we have not performed an observation sensitive to astrophysical BHs below $10\unit{Hz}$, it is difficult to estimate the event rate of binary BHs. However, due to such high horizon redshifts, we would be able to detect several to several tens of the GWs from coalescences of massive BHs if they exist in the Universe.

\section{Conclusions}

We have shown that a neutron DFI has a characteristic frequency sensitive to GWs around $0.1\,{\rm Hz}$ with its size of $\sim 3\unit{km}$ and a GW response of $\Omega^3$ at low frequencies, canceling out all the displacement noises from the mirrors and beam splitters. In contrast to a kilometer-sized laser DFI that has a characteristic frequency, $\sim 10^5\,{\rm Hz}$, a neutron DFI can be utilized for detecting GWs from astrophysical sources below $1\,{\rm Hz}$ that are inaccessible by an ordinary laser interferometer on the ground. Furthermore, the neutron DFI  bridges the frequency bands of LISA and ground-based laser interferometers and plays a crucial role for multi-band observations.

In order to implement a neutron DFI in reality, more studies on experimental aspects and the estimation of sensitivity to a GW are necessary. A simplified detector design for a proof-of-principle experiment of a neutron DFI has been considered in \cite{Iwaguchi:2022hrn}. One of the technical challenges is a reflection angle at a mirror. With the state-of-the art technology, the reflection angle measured from the surface of a mirror is $\sim 5$ degrees at most. However, there is no fundamental limitation on the reflection angle and a large-angle reflection ($\sim 45$-degree) mirror can be realized in principle by a one-dimensional artificial crystal. Additionally, the currently available neutron beams are not very directional to form the geometrical optics of the interferometer and to obtain sufficient intensity~\cite{Rauch-Werner:book}. A more intense pulsed neutron source and an efficient beam transport system have been discussed~\cite{Gunther:2020luk}. Another practical issue to be considered when one demonstrates a neutron DFI in an experiment is imperfections of a detector. The effects on the sensitivity of a neutron DFI from the imbalance of arm lengths, the splitting ratio of beam splitters, the reflectivities of mirrors, and any residuals of displacement-noise cancellation should be evaluated. Therefore, such technical developments are challenging but we believe they are feasible in the future. We leave more detailed investigation on these issues for future work.



\begin{acknowledgments}
A.~N. is supported by JSPS KAKENHI Grant Nos. JP19H01894 and JP20H04726 and by Research Grants from Inamori Foundation. S.~K. is supported by JSPS KAKENHI Grant No. JP19K21875.
\end{acknowledgments}

\bibliography{/Volumes/SSD2/bibliography}
\end{document}